\documentclass[a4paper, 12pt]{article}            
\usepackage{graphicx}  
\begin{document}

\title{\bf Contributions of supernovae type II \& Ib/c to the galactic chemical evolution}
\author{Sandeep Sahijpal}
\maketitle
\begin{center}
Department of Physics, Panjab University, Chandigarh, India 160014;  sandeep@pu.ac.in
\end{center}

{\bf Abstract:} The supernovae SN II \& Ib/c make major stellar nucleosynthetic contributions to the inventories of the stable nuclides during the chemical evolution of the galaxy. A case study is performed here with the help of recently developed numerical simulations of the galactic chemical evolution in the solar neighbourhood to understand the contributions of the SN II and Ib/c by making a comparison of the stellar nucleosynthetic yields obtained by two leading groups in the field. These stellar nucleosynthetic yields differ in terms of the treatment of stellar evolution and nucleosynthesis. The formulation for the galactic chemical evolution is developed for the recently revised solar metallicity of $\sim$0.014. Further, the recent nucleosynthetic yields of the stellar models based on the revised solar metallicity are also used. The analysis suggest that it could be difficult to explain in a self-consistent manner the various features associated with the elemental evolutionary trends over the galactic timescales by any single adopted stellar nucleosynthetic model of SN II \& Ib/c.


\section{Introduction}         

The galactic chemical evolution (GCE) models deal with understanding the origin
and the evolution of the galaxy in context with the evolution of the
isotopic inventories of the stable nuclides (Matteucci \& Franc\c{o}is 1989; Alib\'{e}s et al. 2001; Prantzos \& Aubert 1995; Timmes et al. 1995; Chang et al. 1999, 2002; Goswami \& Prantzos 2000; Sahijpal \& Gupta 2013; Sahijpal 2013). The majority of these works deal with the abundance evolution of the stable nuclides from hydrogen to zinc over the galactic timescales. Subsequent to the big-bang origin of the universe around 13.7 billion years ago, the primordial nucleosynthesis established the initial abundance of hydrogen, helium and lithium in the galaxy. The heavier elements were synthesized by stellar nucleosynthesis within several generations of stars formed over the time-span of $\sim$13 billion years (Alib\'{e}s et al. 2001; Sahijpal \& Gupta 2013).

The stellar nucleosynthetic yields of stars of different masses and
metallicities are some of the most important ingredients of the galactic chemical evolution models (Alib\'{e}s et al. 2001; Sahijpal \& Gupta 2013). The star formation rate, the stellar initial mass function (IMF), the accretion scenario of the galaxy are the other major ingredients to understand the formation and the chemical evolution of the galaxy. The stars of distinct masses and metallicities evolve distinctly and produce a wide range of stable nuclides that contribute to the galactic inventories over distinct timescales. In the present work, an assessment is made regarding the stellar nucleosynthetic contributions of the stars that eventually explode as supernovae SN II \& I b/c to the bulk galactic inventories of the stable nuclides from carbon to zinc. The assessment is based on the nucleosynthetic yields of supernova (SN II \& SN I b/c) resulting from the evolution of ($\geq$11 M$_{\odot}$) stars. The stellar yields of these stars obtained by Woosley \& Weaver (1995) were used in majority of the GCE models developed earlier. Another group has also provided the stellar nucleosynthetic yields of these stars (Chieffi \& Limongi 2004, 2013; Limongi \& Chieffi 2012). Some of the recent models developed by this group have incorporated refinements in the stellar evolution theories by taking into account the stellar rotation and stellar mass loss rates (Chieffi \& Limongi 2013). In this work, we present results of our GCE models based on the stellar yields of these stars obtained by Chieffi \& Limongi (2004, 2013), Limongi \& Chieffi (2012). We also make comparisons of the GCE models based on the stellar nucleosynthetic yields obtained by the two groups (Woosley \& Weaver 1995; Chieffi \& Limongi 2004, 2013) to study the contributions of SN II \& Ib/c to the bulk galactic inventories of the stable nuclides from carbon to zinc.

The majority of the GCE models developed earlier were based on the conventional 
approach of solving the integro-differential equations dealing with the stable 
isotopic abundance evolution of all elements from hydrogen to zinc during the 
galactic evolution (e.g., Matteucci \& Franc\c{o}is 1989;  Prantzos \& Aubert 1995; Timmes et al. 1995; Chang et al. 1999, 2002; Goswami \& Prantzos 2000; Alib\'{e}s et al. 2001). The stellar nucleosynthetic yields of the various stellar phases that include the AGB (asymptotic giant branch) stars, nova and supernovae (SN II, SN Ia and SN I b/c) of different masses and metallicities are incorporated in the equations to consider the temporal isotopic abundance evolution of the galaxy. An alternative numerical approach has been developed recently to simulate the evolution of the galaxy by considering the successive episodes of star formation and their evolution (Sahijpal \& Gupta 2013). In this approach, the galaxy evolves in a realistic manner in terms of formation and evolution of several generations of stars over the galactic timescales (Sahijpal \& Gupta 2013). This recent GCE model incorporates the revised solar metallicity that has been reduced to a value of $\sim$0.014 (Asplund et al. 2009) from an earlier adopted value of $\sim$0.02. The present work is exclusively based on the recently revised solar metallicity. It incorporates for the first time the stellar nucleosynthetic yields of SNII \& Ib/c that are based on the recently revised solar metallicity (Chieffi \& Limongi 2013). It should be noted that the earlier stellar yields of SNII (Woosley \& Weaver 1995; Chieffi \& Limongi 2004) were based on the earlier adopted solar metalicity of $\sim$0.02. Thus, the present work is the first attempt to model GCE with the yields of SNII \& Ib/c based on the revised solar metallicity (Asplund et al. 2009).

\section{Supernova SN II and Ib/c}

The majority of the stable isotopes from carbon to zinc are produced by the stars with the mass $\geq$11 M$_{\odot}$ (e.g., Woosley \& Weaver 1995). The stars in the mass range $\sim$11-30 M$_{\odot}$ eventually explode as core collapse supernova SN II subsequent to their evolution (Sahijpal \& Soni 2006; Sahijpal \& Gupta 2009). This typically occurs rapidly over a timescale of $\leq$25 million years (Myr.) after the formation of the stars within a cluster, thereby, rapidly recycling the stellar nucleosynthetic debris with the interstellar medium. However, the $\geq$30 M$_{\odot}$ stars evolve through Wolf-Rayet stage during which these stars loose significant mass gradually to the interstellar medium (Sahijpal \& Soni 2006; Sahijpal \& Gupta 2009 and references therein). These stars eventually explode as core collapse supernova SN Ib/c over a timescale that is even much shorter than the stars undergoing supernova SN II (Sahijpal \& Gupta 2009). Several factors associated with the stellar evolution of these stars influence the stellar nucleosynthetic yields (Chieffi \& Limongi 2013), and hence, the temporal evolution of the stable isotopic inventories of the various elements from carbon to zinc in the galaxy. The mass-loss rate during the Wolf-Rayet stage is one of the major factors that influence the evolution and the stellar nucleosynthetic yields of these stars (Sahijpal \& Soni 2006; Chieffi \& Limongi 2013). The other major issues that influence the nucleosynthetic yields include the final mass of the stellar remnant at the time of supernova, the incorporation of rotation in the stellar models, the adopted convective criteria and the incorporation of the neutrino induced reactions for the synthesis of nuclides (Woosley \& Weaver 1995; Sahijpal \& Soni 2006; Chieffi \& Limongi 2013).

Woosley \& Weaver (1995) provided the first comprehensive stellar nucleosynthetic models for the $\geq$11 M$_{\odot}$ stars. The stellar yields of the stable and radionuclides from hydrogen to gallium of the stellar models in the mass range 11-40 M$_{\odot}$ with the initial metallicities in the range 0 to $\sim$0.02 were obtained. The upper range of the metallicity refers to the earlier adopted solar metallicity that has been recently revised to a value of $\sim$0.014. These stellar models incorporated the neutrino induced nuclear reactions and the distinct stellar remnant mass cuts for the stars beyond 25 M$_{\odot}$ stars. The stellar yields provided by Woosley \& Weaver (1995) formed the basis for the development of most of the galactic chemical evolution (GCE) models even till now. However, the absence of the stellar nucleosynthetic yields beyond 40 M$_{\odot}$ star was one of the major short-comings associated with the work. In majority of the GCE models, the stellar yields of the stars beyond 40 M$_{\odot}$ stars were obtained by extrapolating the stellar yields of the 11-40 M$_{\odot}$ stars. The uncertainties associated with the extrapolation eventually propagates toward the uncertainties in the GCE models. Further, the influence of the mass-loss rate during the evolution of the stars were not incorporated in the stellar yields obtained by Woosley \& Weaver (1995).

Chieffi \& Limongi (2004) developed an independent set of stellar models for the stars in the mass range of 13-35 M$_{\odot}$ with the initial metallicities in the range 0 to $\sim$0.02. These models provided an alternative set of stellar nucleosynthetic yields to develop GCE models that were earlier based exclusively on the stellar yields obtained by Woosley \& Weaver (1995). It was demonstrated by Chieffi \& Limongi (2004) that the stellar yields of the stars with low metallicity ($\leq$10$^{-4}$) does not depend upon the initial stellar composition. As a part of the present work, we make comparisons of the GCE models developed based on the stellar yields obtained by Woosley \& Weaver (1995) and Chieffi \& Limongi (2004). The stellar models developed by Chieffi \& Limongi (2004) had limitations identical to that of the models developed by Woosley \& Weaver (1995) in terms of the limited stellar mass range and the exclusion of the mass-loss rates during the stellar evolution of the stars.

Limongi \& Chieffi (2012) have recently developed for the first time the stellar models for the stars in a wide mass range of 13-80 M$_{\odot}$ with the initial zero metallicity. Further, Chieffi \& Limongi (2013) have developed the stellar models for the stars in a wide mass range of 13-120 M$_{\odot}$ with the revised solar metallicity of $\sim$0.014. The recent stellar models incorporate stellar rotation and mass-loss rates. In the present work, we have developed GCE models based on the stellar nucleosynthetic yields of the stars obtained earlier by Woosley \& Weaver (1995), and recently by Chieffi \& Limongi (2004, 2013) and Limongi \& Chieffi (2012). In comparison to the limited mass range of 11-40 M$_{\odot}$  (Woosley \& Weaver 1995), the recent works cover a wide range of 11-100 M$_{\odot}$ in terms of the stellar nucleosynthetic yields. This reduces the uncertainties associated with the extrapolation of nucleosynthetic yields for greater than 40 M$_{\odot}$  stars. The entire discussion of the present work is exclusively based on the available stellar nucleosynthetic yields of these two groups. We have not made any further attempt to understand the complexities associated with the stellar evolution and its influence on the nucleosynthesis (see e.g., Georgy et al. 2013). Except for SN Ia, we have also avoided the possibilities of massive binary systems exploding as supernova and hypernova (see e.g., Sahijpal \& Soni 2006)

\section{Numerical simulations}

The simulations of the GCE were performed assuming a gradual accretion of the galaxy in two episodes (Chiappini et al. 1997). The initial accretionary phase resulted in the growth of the galactic halo and the thick disk over an assumed timescale of one billion years. The second episode followed the gradual accretional growth of the galactic thin disk over a timescale of around seven billion years. It has been observered that this adopted criteria of the accretion of galaxy is able to explain majority of the trends in the abundance evolution of the elements. Further, in order to avoid the G-dwarf metallicity problem, the metallicity of the accreting matter on the galaxy was assumed to be 0.1 times the final acquired metallicity of the sun (e.g., Alib\'{e}s et al. 2001; Sahijpal \& Gupta 2013).

We simulated the evolution of the galaxy around the present position of the sun that is defined as the solar neighbourhood. The solar neighbourhood was confined to an annular ring within 7 to 8.9 kiloparsecs from the center of the galaxy (Sahijpal \& Gupta 2013). The simulations were performed with a time-step of one million years (Myr.) that is less than timescale of $\sim$ 3.5 Myr. for the evolution of the most massive star formed within the simulation. The simulations were executed by the synthesis of stellar populations of sucessive generations from the interstellar medium of an evolving metallicity. The simulated stars were synthesized and evolved according to their masses and metallicities. The nucleosynthetic contributions of the evolved stars in the form of stellar ejecta were incorporate into the evolving interstellar medium (Sahijpal \& Gupta 2013). The successive generations of stars were formed within the solar annular ring on the basis of an assumed star formation rate that depends upon the prevailing interstellar gas surface mass density and the total matter surface mass density of the ring. We have made use of the star formation rate formulation developed by Alib\'{e}s et al. (2001). This formulation is based on the star formation rate proposed by Talbot \& Arnett (1975) and Dopita \& Ryder (1994). Based on our recent works on the influence of the galactic chemical evolution on the star formation rate in the earliest one billion years of the evolution of the galaxy that corresponds to the accretion era of the galactic halo and the thick disk (Sahijpal 2012, 2013), we have assumed a general enhancement in the star formation rate in the era by a factor of three. This correspond to the use of a value of 3 for $\nu$ in the star formation rate (Alib\'{e}s et al. 2001) during the initial one billion years and a value of 1 for $\nu$ for the subsequent time in the evolution of the galaxy. The essential trends in the elemental evolution, specifically, in the case of oxygen and iron can be explained in a better manner with this choice of star formation rate parameters (Sahijpal 2012, 2013).

A normalized three stage stellar initial mass distribution function (IMF), $\phi$(m) = A m$^{-(1+x)}$, was assumed in the mass range 0.1-100 M$_{\odot}$ to synthesize the simulated stars at the time of star formation (Matteucci 2003) based on the criteria used by Sahijpal \& Gupta (2013). According to the adopted criteria, the IMF differential spectra consists of stars with the masses represented by the integer numbers in the mass range 3-100 M$_{\odot}$. However, in the low mass range, we choose the stellar masses of 0.1, 0.4, 0.8, 1.0, 1.25, 1.75 and 2.5 M$_{\odot}$ in order to cover the critically important mass range in an appropriate manner. This choice was also partially motivated by the availability of the stellar nucleosynthetic yields of the AGB stars corresponding to some specific masses (Karakas \& Lattanzio 2007). The power index, {\it x}, was assumed to be 0 and 1.7 in the mass ranges, 0.1-1 M$_{\odot}$ and 1-8 M$_{\odot}$, respectively. This is almost consistent with the earlier works regarding the choice of the IMF parameter values in this mass range (Matteucci 2003; Sahijpal \& Gupta 2013). It should be mentioned that there have been several proposals for values of the power index for the IMF (Matteucci 2003). The normalizing constant, {\it A}, was chosen to synthesize the IMF according to the star formation rate prevailing at any epock. This constant will determine the total number of stars formed corresponding to different masses at any specific time with a uniquely defined metallicity that will depend upon the metallicity of the interstellar medium prevailing at that time. The value of the power index in the stellar mass range 11-100 M$_{\odot}$ was treated as one of the simulation free parameters to obtain the value of $\sim$0.014 for the solar metallicity at the time of formation of the solar system around 4.56 billion years ago. Even though, the power index in the stellar mass range 11-100 M$_{\odot}$ is treated as a free parameter, it was observed that in all the simulations it adopted a value within the range predicted in the earlier works (Matteucci 2003). The detailed parametric analyses regarding the various parameters of the simulations have been performed earlier (Sahijpal \& Gupta 2013; Sahijpal 2013). In the present work, we choose only a selective range of the parameters to understand the role of the massive stars in the galactic chemical evolution.

The stars of different masses and metallicities evolve distinctly over different timespans. We have considered the stellar nucleosynthetic contributions of the AGB stars, supernovae SN Ia, SN II and SN Ib/c to the evolving interstellar medium. The detailed mass balance calculations were performed to incorporate the stellar contributions to the interstellar medium subsequent to the evolution of distinct populations of stars.

\subsection{Nucleosynthetic contributions of AGB stars and supernova SN Ia}

The stars in the mass range 0.1-8 M$_{\odot}$ evolve through AGB phase (Karakas \& Lattanzio 2007). These stars produce the intermediate mass nuclei along with the s-process and the neutron-rich nuclei. We have considered the stellar nucleosynthetic yields of the AGB stars in the mass range 1.25-8 M$_{\odot}$ for the wide range of stellar metallicities (Karakas \& Lattanzio 2007). As mentioned earlier, the choice of the stellar IMF in the low and intermediate mass range was partially motivated by the availability of the stellar yields for some specific masses.

The supernova SN Ia nucleosynthetic yields obtained by Iwamoto et al. (1999) for the various stellar models were used. The SN Ia rate adopted in the present work is based on the normalized time delay distribution function for the SN Ia rate that has been adopted by Matteucci et al. (2009) to trigger the SN Ia. This normalized distribution function is based on the essential requirement of an initiation of the rapid SN Ia rich nucleosynthetic contributions within the timescales of $\sim$100 million years (Myr.) from the time of formation of the binary systems that eventually produce SN Ia. Along with an enhanced star formation rate in the initial one billion years of the formation of the galaxy, the adopted prompt SN Ia contribution explain majority of the galactic evolutionary trends in the elemental abundance distribution of the solar neighbourhood (Sahijpal 2013). A fraction, {\it f}, of the stars produced during any single episode of star formation was treated as binary star pairs that would eventually undergo SN Ia (Sahijpal \& Gupta 2013). The parameter, {\it f}, along with the power index, {\it x}, corresponding to the stellar initial mass function in the stellar mass range 11-100 M$_{\odot}$ were considered as the free parameters to reproduce the solar metallicity of $\sim$0.014 and [Fe/H] = 0 at the time of formation of the solar system around 4.56 billion years ago (Sahijpal \& Gupta 2013). It should be mentioned here that in case we fix the value of the power index {\it x}, corresponding to the stellar IMF in the stellar mass range 11-100 M$_{\odot}$ on the basis of its value used in the literature (see e.g., Matteucci 2003), we will end up in a distinct solar metallicity at the time of formation of the solar system. Since, the composition of the sun is one of the most important ingredients to understand the galactic chemical evolution, we adopted the criteria for parameterizing the power index. However, it should be noted that values obtained for the parameter in the various simulations are well within the range predicted earlier (Matteucci 2003).

\subsection{Nucleosynthetic contributions of SN II \& Ib/c}

As mentioned earlier, we have made comparisons of the GCE models based on the stellar nucleosynthetic yields of the SN II \& Ib/c obtained by Woosley \& Weaver (1995), Chieffi \& Limongi (2004, 2013) and Limongi \& Chieffi (2012). We ran three simulations in order to make the comparisons. The GCE model, {\bf WW95}, is based on the stellar yields obtained by Woosley \& Weaver (1995). The stellar yields obtained by Chieffi \& Limongi (2004) were used in the GCE model, {\bf CL04}. In the GCE model, {\bf CL13}, the stellar yields of the zero metallicity stars in the mass range 13-80 M$_{\odot}$, obtained recently by Limongi \& Chieffi (2012) were used. Further, the stellar yields of the revised solar metallicity of $\sim$0.014 in the stellar mass range 13-120 M$_{\odot}$ (Chieffi \& Limongi 2013) were used in this simulation. These stellar models incorporate stellar rotation and mass-loss. Due to the absence of the stellar yields corresponding to the intermediate metallicities, the stellar yields corresponding to the available intermediate stellar metallicities were taken from the earlier work of the same group (Chieffi \& Limongi 2004) in the the {\bf CL13} GCE model. This will remain one of the limitations of the present approach until the stellar yields corresponding to the intermediate metallicities become available in literature. We have addressed the associated repercussions in the following. The stellar yields of the remaining intermediate stellar masses and metallicities were interpolated appropriately during the simulation (Sahijpal \& Gupta 2013). Due to the uncertainties in the mass fall-back at the time of supernovae SN II and SN Ib/c, the iron yields of all the stars beyond 30 M$_{\odot}$ were systematically reduced by a factor of two as a standard practice (e.g., Timmes et al. 1995; Sahijpal \& Gupta 2013).

\section{Results and Discussion}

A comparison of the GCE models based on the distinct set of nucleosynthetic yields of SNII \& Ib/c is performed in the present work to understand the influence of the galactic chemical evolution on the nucleosynthetic yields of SNII \& Ib/c. The GCE models have been developed based on the recently adopted approach of numerically simulating the evolution of the galaxy in terms of formation of successive generations of stars. Further, the GCE models have been developed to reproduce the recently revised solar metallicity of $\sim$0.014 at the time of formation of the solar system around 4.56 billion years ago. As mentioned earlier, this was achieved by fitting the parameters, {\it x} and {\it f}, related with the stellar initial mass function power index in the mass range 11-100 M$_{\odot}$ and the fraction of the stars produced as binary stars at the time of star formation, respectively. The assumed binary systems will eventually evolve to SN Ia. We obtained the values of $\sim$0.042 and $\sim$1.51 for {\it f} and {\it x}, respectively, in the case of model WW95. In case of the model CL04, we deduced these values to be $\sim$0.053 and $\sim$1.57, respectively, whereas, the model CL13 inferred the value of $\sim$0.054 and $\sim$1.64, respectively. The deduced power exponents are well within the range anticipated for the IMF in the earlier works (Matteucci 2003). As mentioned earlier, the compulsion to reproduce the solar metallicity at the time of formation of the solar system necessitate the role of the power index as one of the critical free parameters. The CL13 GCE model infers steeper slope for the IMF in the mass range 11-100 M$_{\odot}$ compared to the GCE model CL04 which in turn infers steeper slope compared to the GCE model WW95. Further, the GCE WW95 model infers the least SN Ia contributions, whereas, the model CL13 infers the highest SN Ia contributions.

The predicted star formation rates for the three simulations corresponding to the distinct supernova nucleosynthetic yields are presented in Fig. 1. As discussed earlier, the enhanced star formation rate in the initial one billion years of the evolution of the galaxy that is represented by a choice of the value of 3 for $\nu$ in the star formation rate (Alib\'{e}s et al. 2001) explains the essential features of the elemental evolution trends (Sahijpal 2012, 2013). The differences in the star formation rates for the three simulations originate due to the differences in the SN II \& Ib/c stellar contributions of the major elements, e.g., carbon, oxygen, etc. towards the bulk inventories of the interstellar medium as is elaborately discussed in the following.

The predicted evolution of the metallicity of the solar neighbourhood for the three models, WW95, CL04 and CL13, over the galactic timescales are presented in Fig. 2. It should be noted that the WW95 model based on the nucleosynthetic yields obtained by Woosley \& Weaver (1995) yield almost identical evolutionary trend as compared to the model CL04 based on the yields obtained by Chieffi \& Limongi (2004). However, the evolutionary trend for the CL13 model is quite distinct compared to the other two models. The CL13 model differs from the CL04 model due to the incorporation of the recent nucleosynthetic yields of the stellar models based on the revised solar metallicity of $\sim$0.014 along with the inclusion of stellar rotation and mass-loss rates during the evolution of the stars (Chieffi \& Limongi 2013). Further, the stellar nucleosynthetic yields of the zero metallicity stars (Limongi \& Chieffi 2012) have been used in CL13 model instead of the earlier used stellar yields of the zero metallicity models (Chieffi \& Limongi 2004).

The oxygen, followed by carbon, are the main contributors to the metallicity of the evolving galaxy. The steeper slope in the metallicity rise of the CL13 model compared to the other models can be attributed to the higher carbon and oxygen yields of the stellar models by Chieffi and Limongi (2013) compared to the earlier stellar models (Woosley \& Weaver 1995; Chieffi \& Limongi 2004) corresponding to the solar metallicity. As mentioned earlier, the GCE model CL13 infers the steepest stellar initial mass function among all the models. Thus, the higher production of oxygen and carbon by SN II \& SN Ib/c in the CL13 GCE model would necessitate a slow rise in the metallicity during the initial epoch of the galaxy. This will gradually rise to the solar metallicity value around 4.56 billion years ago. However, it should be noted that in the CL13 model due to the absence of the stellar yields for the stellar models with less than the solar metallicity we have used the corresponding yields obtained earlier by Chieffi \& Limongi (2004) that excludes the effects of stellar rotation and mass-loss. It is anticipated that the stellar yields of at least the primary nuclides, e.g., $^{12}$C and $^{16}$O that does not substantially change with the change in the stellar metallicity, the incorporation of stellar rotation will result in a further enhancement in the stellar yields of these nuclides for the stellar models with less than the solar metallicity. This will further alter the evolutionary trend in the metallicity rise of the CL13 model. In order to make a direct comparison of the stellar nucleosynthetic yields of the three stellar models, the net accumulative stellar yields (in M$_{\odot}$) of the various stable nuclides from the supernovae SN II \& Ib/c of the stars from a single stellar population that could represent either a single stellar cluster or associated stellar clusters formed together at a specific time in the simulations with the defined stellar IMF estimated at metallicities of 0.0014 (0.1 $\times$ Z$_{\odot}$) and 0.014 (Z$_{\odot}$) for the three models are presented in Fig. 3. These yields were numerically obtained for each stable isotope by multiplying the stellar yields of the massive stars with mass greater than 11 M$_{\odot}$ with the stellar IMF value for the specific stellar mass prevailling at the specific time corresponding to the metallicities of 0.0014 (0.1 $\times$ Z$_{\odot}$) and 0.014 (Z$_{\odot}$) of the interstellar medium. The net accumulative yields were obtained by adding the stellar yields corresponding to the various stellar masses. It should be noted that the star formation rate prevailling at the specific time will determine the total mass of the stellar population.

The evolutionary trends for the [Fe/H] is presented in Fig. 4 for the three GCE models. All the GCE models are able to explain the observational spread in the [Fe/H] value of the F, G and K dwarf stars in the solar neighbourhood. It should be noted that the GCE trends are exactly fitted to explain the [Fe/H] = 0 at the time of formation of the solar system around 4.56 billion years ago rather than average of the exact observational spread around 4.56 billion years ago. The differences in the [Fe/H] evolutionary trends between the WW95 model and the other two models, namely, CL04 and CL13 is due to the lower iron yields (Fig. 3) of the stars by Chieffi \& Limongi (2004, 2013) compared to that obtained by Woosley \& Weaver (1995). Since, SN Ia is the major source of iron ({\it e.g.}, Alib\'{e}s et al. 2001; Sahijpal \& Gupta 2013; Sahijpal 2013), the lower iron yields of the SNII \& Ib/c in CL04 and CL13 models compared to the model WW95 (Fig. 3) is compensated by a higher rates of SN Ia. As mentioned earlier, this is reflected in the higher anticipated fraction, {\it f}, of the binary stellar systems formed that eventually explode as SN Ia in the CL04 and CL13 models compared to the model WW95.

The evolutionary trends in the metallicity (Fig. 2) and [Fe/H] (Fig. 4) for the three GCE models also influence the anticipated supernovae history of the galaxy apart from the predicted star formation rates (Fig. 1). The predicted trends in the SN II + I b/c and SN Ia rates are presented in Fig. 5. As mentioned earlier, the oxygen and carbon yields of the recent stellar models (Chieffi \& Limongi 2004, 2013) are higher than the earlier models (Woosley \& Weaver 1995), whereas, the iron yields of the recent models are low corresponding to the solar metallicity (Fig. 3). In comparison with the WW95 GCE model, this results in a reduction in the SN II + I b/c rates in the CL04 and CL13 GCE models with an increase in the SN Ia rates. This also influences the predicted surface mass density of the stellar remnants of SN II + I b/c, i.e., the neutron stars and black holes (Fig. 6) that shows a corresponding reduction in the surface mass densities of the models CL04 and CL13 with respect to the model WW95. Apart from the surface mass densities of the stellar remnants, the predicted total surface mass density, the stellar mass density and the gas mass density are also presented in Fig. 6. The predicted total surface mass density of the present epock was fitted to explain the present observed value of $\sim$ 54 M$_{\odot}$ pc$^{-2}$ (Alib\'{e}s et al. 2001) in all the simulations. Since, a case study of the stellar contributions from the massive stars is presented in the present work, the white dwarfs that are the remnants of red-giant and AGB stars exhibit identical trands in their predicted surface mass densities for the various models.

The normalized stable isotopic yields obtained from the three GCE models are presented in Fig. 7 at the time of formation of the solar system around 4.56 billion years ago. It should be noted that due to the uncertainties associated with the various physical processes associated with the galactic and stellar evolution, a variation by a factor of two over the anticipated normalized value is considered to be tolerable. The three GCE models based on the distinct set of stellar yields predict almost identical stable isotopic abundances for the elements lighter than sulphur within a factor of two. An identical trend continues even in the case of elements from titanium to nickel. However, the GCE predictions for the three models are significantly different in the case of fluorine, chlorine, argon, potassium, copper, zinc and gallium, with the recent stellar nucleosynthetic models (Chieffi \& Limongi 2004, 2013) inferring lesser stellar yields compared to the earlier models (Woosley \& Weaver 1995).

The results obtained from the normalized elemental abundance evolution of twelve elements over the galactic timescales are presented in Fig. 8. Compared to the WW95 GCE model, the galactic evolutionary trends of the CL04 GCE model infer better match for carbon, nitrogen, oxygen, magnesium, cobalt and nickel. The galactic evolutionary trend for nickel becomes even better in the CL13 GCE model. The normalized trends for the CL04 and CL13 GCE models are almost identical for nitrogen, magnesium, aluminium, silicon, sulphur and calcium. The galactic evolution of nickel and zinc are best explained by the CL13 model. However, it is not possible to explain the galactic chemical evolution of all the elements consistently by any single adopted nucleosynthetic model.

\section{Conclusions}

Numerical simulations of the galactic chemical evolution have been performed in the present work to understand the contributions of supernovae SN II \& Ib/c. A comparison of the stellar nucleosynthetic yields of the stars (11-100 M$_{\odot}$) obtained by distinct groups is performed based on these numerical simulations. Some of these adopted recent stellar models incorporate stellar rotation and mass-loss rates. Based on the present analysis it seems that it could be difficult to explain all the features of the galactic chemical evolution of all the elements from hydrogen to zinc in a self-consistent manner by any single adopted stellar nucleosynthetic model of the supernovae SN II \& Ib/c. 

\vspace{2cm}
{\bf Acknowledgements:}

We are extremely grateful to the numerous comments made by the reviewer that led to significant improvement of the manuscript. This work is supported by the PLANEX (ISRO) grant

\begin{figure}
\includegraphics[width=1.0\textwidth]{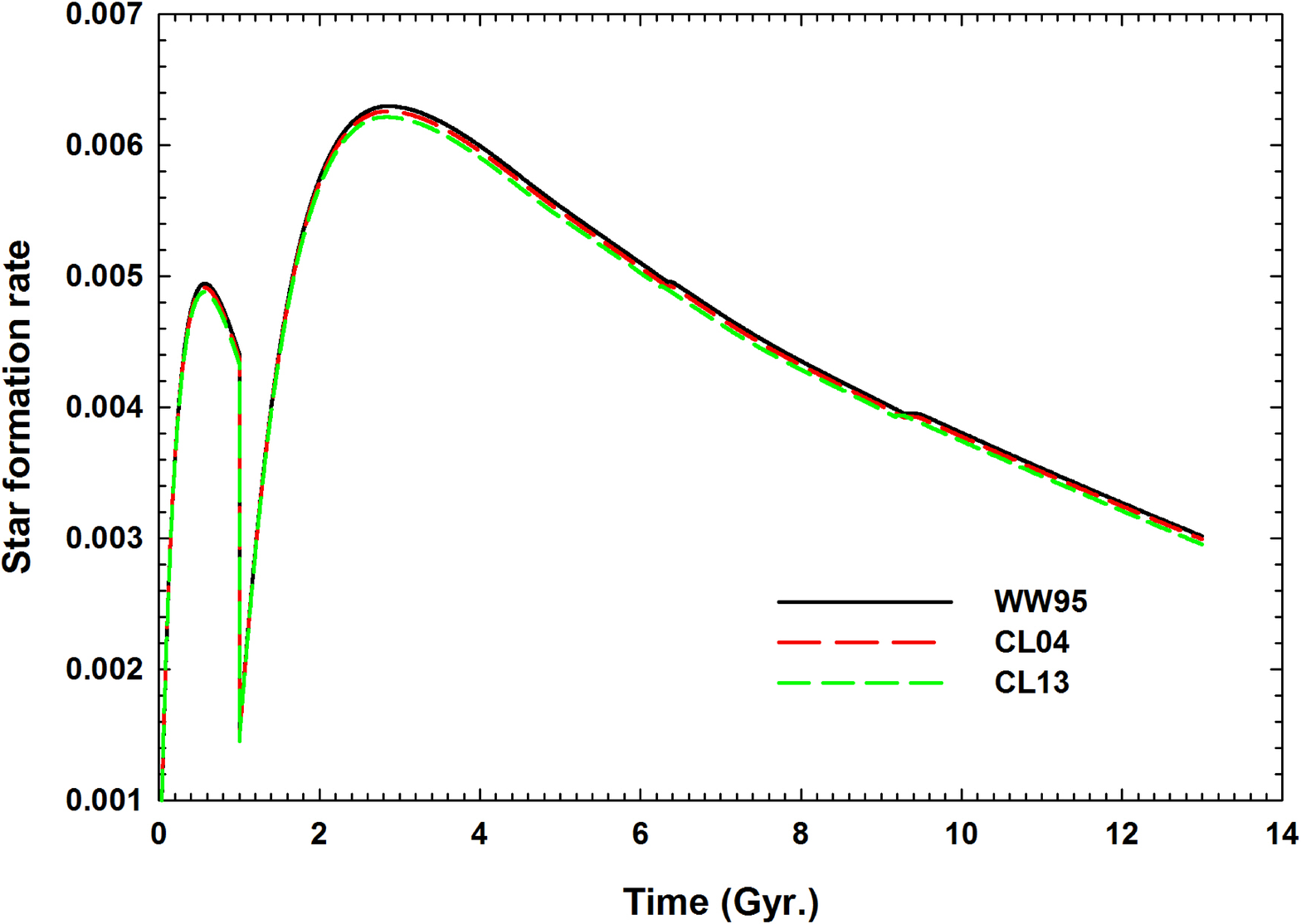}
\caption{The star formation rate (M$_{\odot}$ pc$^{-2}$ Myr.$^{-1}$) for the three distinct GCE models, WW95, CL04, CL13, with distinct stellar nucleosynthetic yields.}
\end{figure}

\begin{figure}
\includegraphics[width=1.0\textwidth]{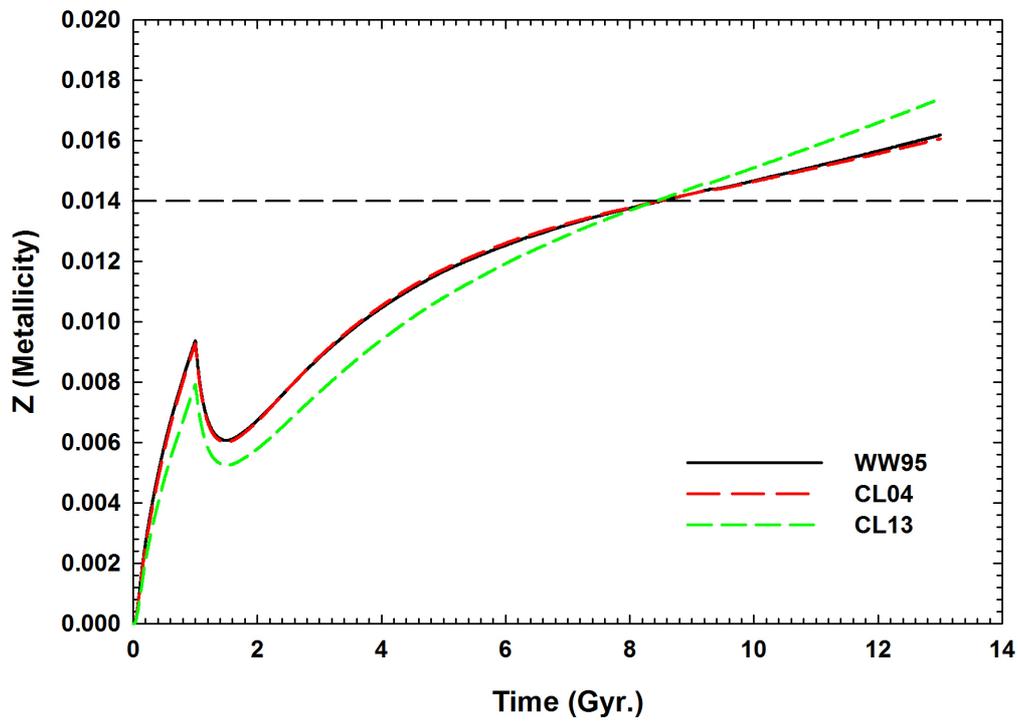}
\caption{The predicted evolution of the metallicity for the three distinct GCE models, WW95, CL04, CL13, with distinct stellar nucleosynthetic yields. The revised solar metallicity is presented as a horizontal dashed line.}
\end{figure}

\begin{figure}
\includegraphics[angle=90, width=0.7\textwidth]{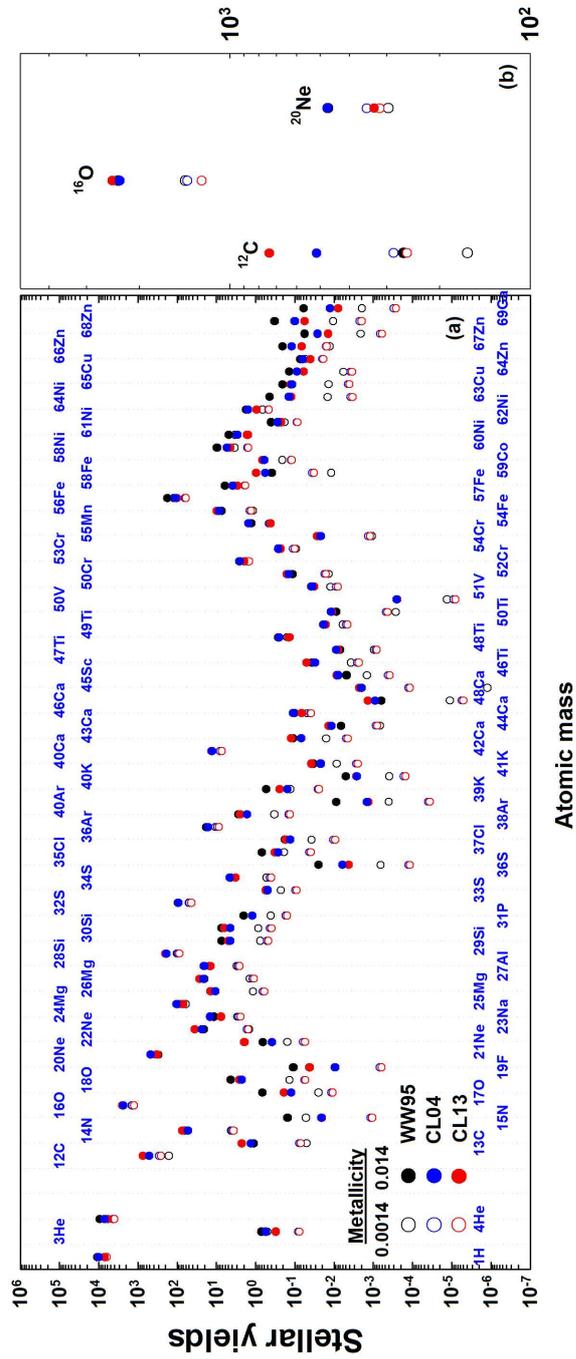}
\caption{The accumulative stellar yields (in M$_{\odot}$) of the various nuclides from the supernovae SN II \& Ib/c of the stars from a single stellar population formed at a specific time with a defined metallicity and IMF.}
\end{figure}

\begin{figure}
\includegraphics[width=1.0\textwidth]{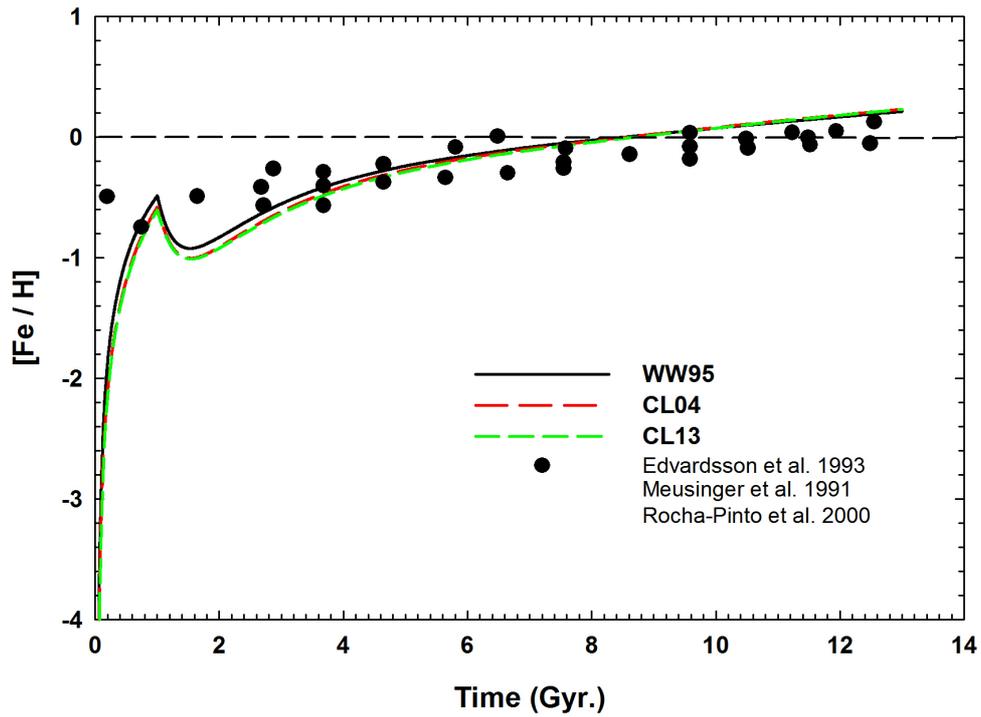}
\caption{The predicted evolution of [Fe/H] for the three GCE models with distinct set of stellar nucleosynthetic yields. The solar value is represented by dashed line. The observational spread of the F, G and K dwarf stars in the solar neighbourhood obtained by three groups is presented for comparison (Alib\'{e}s et al. 2001)}
\end{figure}

\begin{figure}
\includegraphics[width=1.0\textwidth]{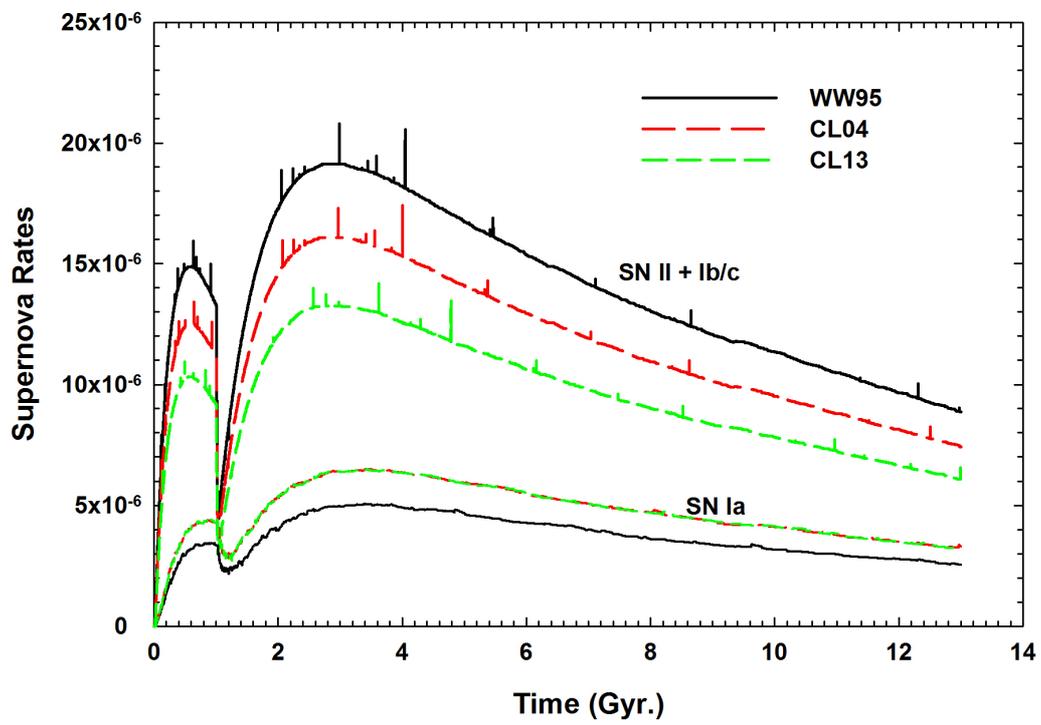}
\caption{The predicted supernova rates (in pc$^{-2}$ Myr.$^{-1}$) for the three GCE models.}
\end{figure}

\begin{figure}
\includegraphics[width=1.0\textwidth]{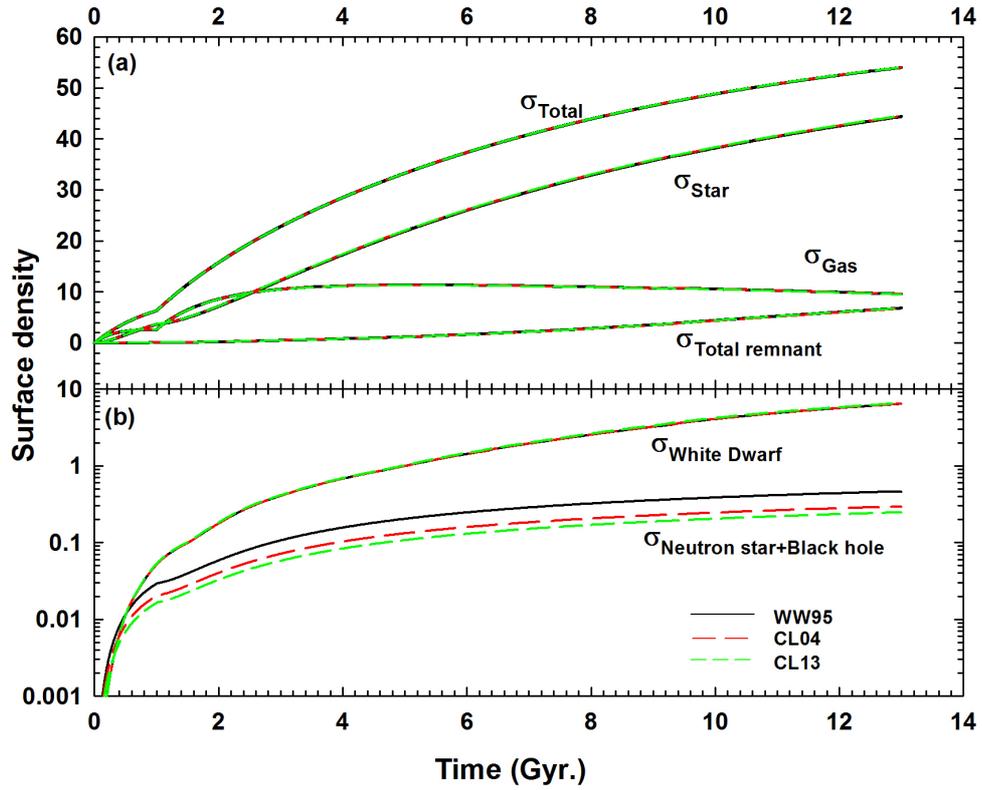}
\caption{(a) The predicted surface mass density (M$_{\odot}$ pc$^{-2}$) of the total surface mass distribution, the stellar mass density, the gas mass density and the stellar remnant mass densities for the three GCE models. (b) The predicted surface mass density of the stellar remnants that include white dwarfs, neutron stars and black holes}
\end{figure}

\begin{figure}
\includegraphics[width=1.0\textwidth]{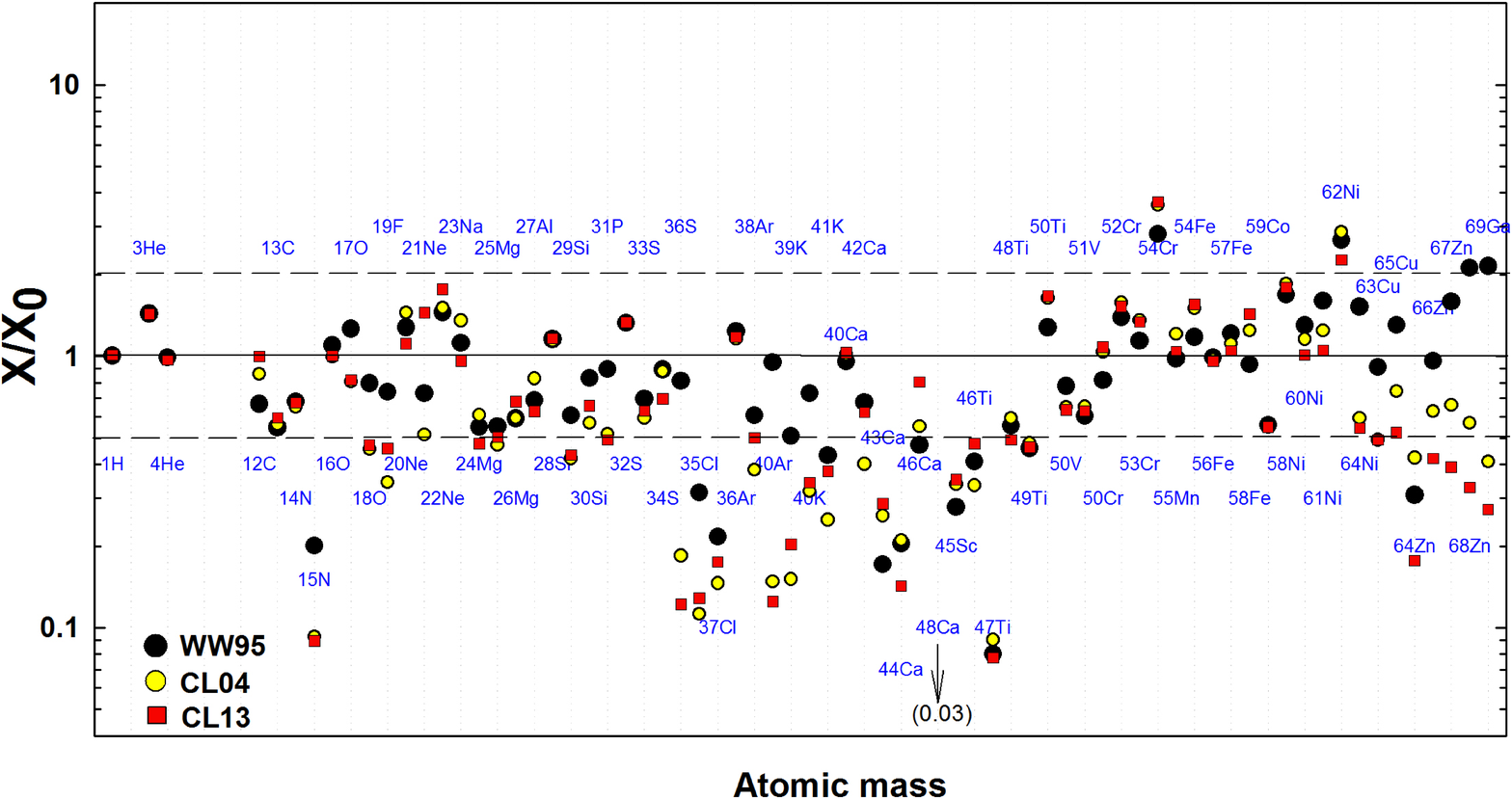}
\caption{The predicted normalized isotopic abundances for the three GCE models with distinct set of stellar nucleosynthetic yields. The abundances were estimated at the time of formation of the solar system around 4.56 Gyr. The revised solar metallicity was assumed to be 0.014}
\end{figure}

\begin{figure}
\includegraphics[width=1.0\textwidth]{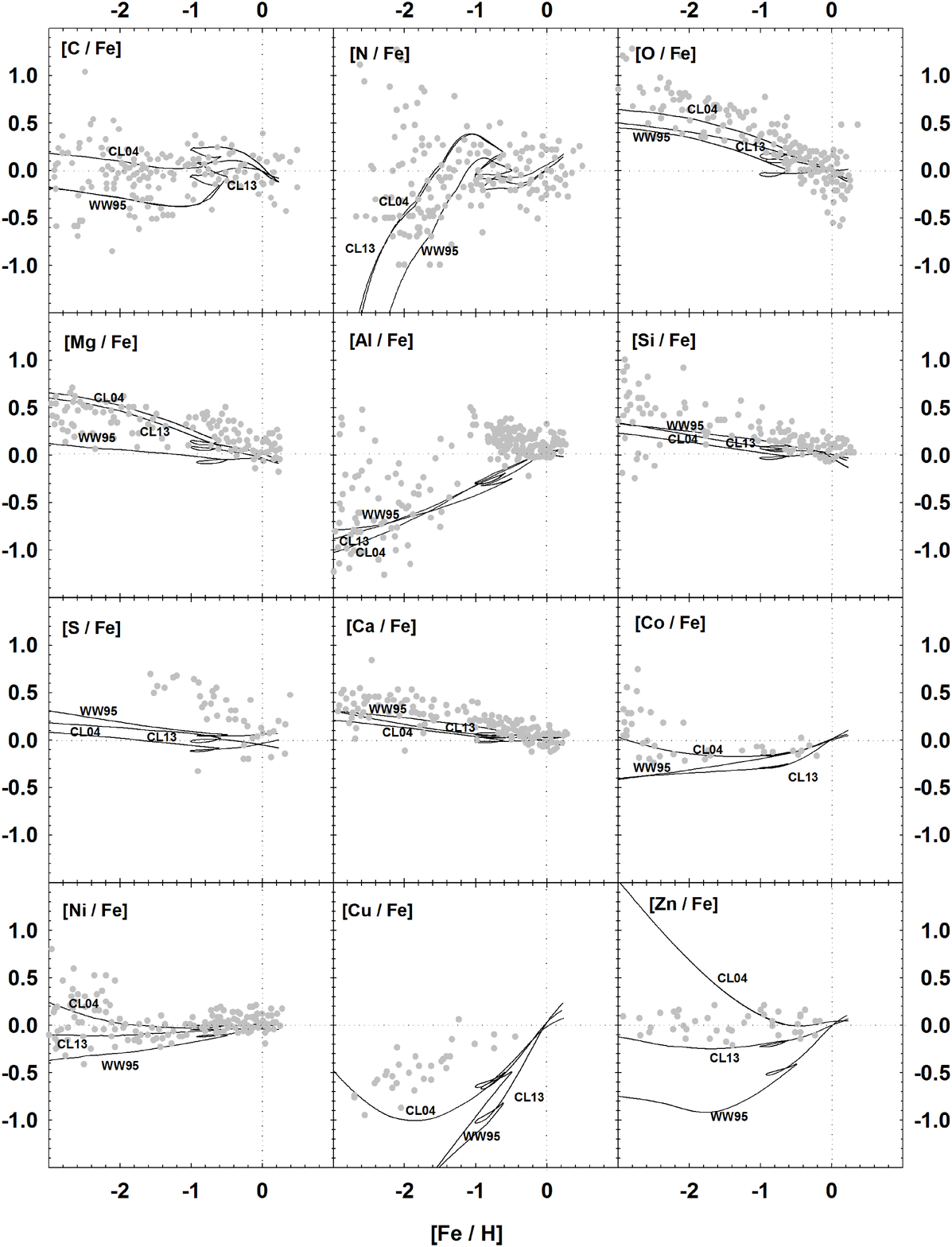}
\caption{The predicted normalized elemental abundance evolution for the various models corresponding to different set of stellar yields. The observational data of the F, G and K dwarf stars in the solar neighbourhood have been adopted for comparison. The detailed references can be found in the recent work by Sahijpal \& Gupta (2013)}
\end{figure}

\end{document}